# From ontology design to user-centred interfaces for music heritage

Giulia Renda[1], Marco Grasso[2], Marilena Daquino[3]

[1]University of Bologna, Italy - giulia.renda3@unibo.it
[2]University of Bologna, Italy - marco.grasso7@unibo.it
[3]University of Bologna, Italy - marilena.daquino2@unibo.it

## ABSTRACT

In this article we investigate the bridge between ontology design and UI/UX design methodologies to assist designers in prototyping web applications for information seeking purposes. We briefly review the state of the art in ontology design and UI/UX methodologies, then we illustrate our approach applied to a case study in the music heritage domain.

## KEYWORDS

Music heritage; ontology design; ux design; generous interfaces

## 1. INTRODUCTION

Polifonia[1] is a European project that aims at connecting resources in the music heritage and to engage with experts and the general public. Ten pilot projects have been designed for the purpose, wherein scholars collect data sources (e.g. texts, audio files), extract information, and transform data into Linked Open Data (LOD) to populate a knowledge graph. The knowledge extraction is driven by competency questions, which also guide the ontology design. The ultimate goal of the knowledge graph is to be leveraged in one or more web applications.

However, eliciting data/user requirements, ensuring completeness, and selecting the right approach – e.g. modular, bottom-up, user-centred – to develop User Interfaces (UI) and Experience (UX) based on domain requirements is challenging. While several efforts have been made to integrate methods of Human-Computer Interaction (HCI) into ontology design methodologies, to the best of our knowledge, there is no overall ontology-driven methodology for developing multi-purpose web solutions. In particular, how can we create interfaces for presenting data characterised by a broad and diverse scope leveraging Linked Open Data and domain ontologies? How do we ensure interfaces are specialised enough to answer complex questions but are usable by stakeholders with different backgrounds?

In this article we investigate the bridge between ontology design and UI/UX design methodologies to assist designers in prototyping web applications for information seeking purposes. We briefly review the state of the art in ontology design and UI/UX methodologies, then we illustrate our approach applied to the Polifonia case study. The methodology can be reused in similar contexts where cultural heritage data are disseminated on the web.

## 2. STATE OF THE ART

In the Semantic Web, the structure of SPARQL queries is closely related to ontologies. This means that users' information needs can be mapped to both ontology requirements and interfaces requirements. Therefore, we would expect that ontology design methods are closely related to UI/UX methodologies. Several ontology design methodologies have indeed adopted tools from HCI. Such methods favour a bottom-up approach to elicit requirements and rely on the intervention of domain experts in (1) defining the domain space and vocabulary, (2) outline motivating scenarios, and (3) extracting requirements in the form of natural language Competency Questions (CQs) [8] [14] [21] [5].

The eXtreme Design (XD) methodology [16] [3] prescribes practices for capturing goals, interests, and tasks from stakeholders, grouping them under umbrella categories, i.e. personas [10]. Unlike other methodologies, XD also captures research journeys, expectations, and priority levels. Since personas also drive the definition of users' behaviours search interfaces, the knowledge acquisition process is closer to current practices in UI/UX design. For these reasons, XD is a good candidate for a seamless integration into UI/UX design processes.

When mapped to UI elements, interactive behaviours can lead to two types of search paths. On the one hand, exploratory journeys can focus on retaining the user, breaking down the information in small chunks to reduce the cognitive load

---

[1] https://polifonia-project.eu/

[11]. On the other hand, preventing the user from seeing the "whole picture" could be disorienting. Studies in Information Science suggest that "third generation" information systems should first filter out data of interest and then apply data analysis and knowledge discovery tools on the target [9]. To meet this call, scholars advocate for more generous interfaces [22], leading to an approach based on "overview first, zoom and filter, then details on demand" [19].

To reconcile these two different perspectives, the user interface must be able to support multiple tasks and user journeys [1]. In this respect, stories are powerful tools for designing experiences, as they present facts connected by causal relationships, and help to formulate users' motivational aspects or to describe unforeseen situations [7].

Design Thinking (DT) [18] [2] [6] [13] is a user-centred approach to problem solving, based on a hypothesis-driven, abductive and dialectical approach to map requirements to design ideas. Previous studies have shown that DT effectively improves the quality of the ideas generated and reduces the risk of failure [12]. It consists of six phases: empathise, define, ideate, prototype, test, and implement. In the data collection phase, various methods are used, including personas, stories, stakeholder and user journey maps [4]. Several studies [17] [20] [15] have attempted to incorporate TD into specific aspects of ontologies creation. Results mostly provide formal definitions of HCI and DT methods, but do not inform us on how to leverage real-world domain ontologies in the DT process. In this work, we aim at filling this gap, suggesting the application of methods and analyses widely recognised as tools of the DT methodology directly to the domain ontologies during the ontology design phase.

## 3. METHODOLOGY

We introduce a modular workflow harmonising eXtreme Design and Design Thinking, that reuses content/user requirements in UI/UX design. We identify nine stages, grouped in three main activities (Figure 1), namely:
- **Ontology design.** (1) The ontology design team outlines personas and groups them, (2) writes one or more stories for each persona, and (3) extracts competency questions.
- **User interfaces design.** (4) The web development team (us) defines the most important CQs, which serve as drivers for all others, and (5) group remaining questions into meaningful clusters. (6) We analyse drivers and clusters and select appropriate visualisation types for the reference data.
- **User experience design.** (7) The web development team outlines interaction patterns (via competitive analysis or focus groups), (8) selects appropriate solutions to deploy, and (9) performs user testing validation.

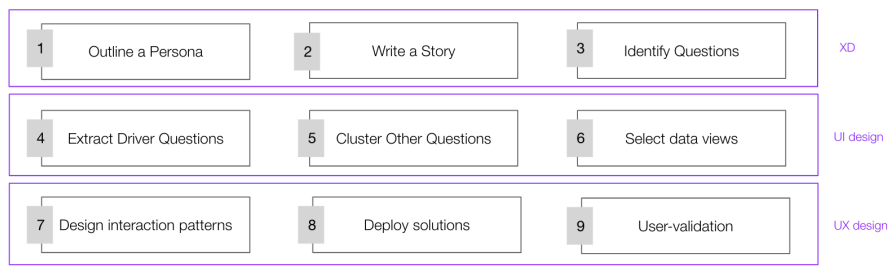

**Figure 1. Overview of the methodology**

## 4. CASE STUDY: DESIGNING POLIFONIA INTERFACES

The Polifonia ontology network describes music sources, performances, instruments, and music features. The Polifonia ecosystem currently includes 9 datasets, for which 19 personas, 28 stories, and 240 competency questions have been identified[2]. Following our workflow, we have identified four web applications to be developed, namely: musoW[3], a filter-based catalogue of music data on the web, targeted to music professionals; MELODY[4], a web editor of data visualisations and stories, targeted to music domain experts; Corpus[5], to perform linguistic analysis over a vast corpus of music-related text sources; and the Polifonia Web portal (in progress), to present data to lay users according to several strategies. An example of the workflow applied to Polifonia for generating musoW is the following.
1. **Persona building.** Laurent is a music journalist. He searches the web for new music sources, but he does not have sufficient technical skills to perform these searches systematically.

---
[2] https://github.com/polifonia-project/stories
[3] https://projects.dharc.unibo.it/musow/
[4] https://projects.dharc.unibo.it/melody/
[5] https://polifonia.disi.unibo.it/corpus/

2. **User Story.** Laurent publishes a weekly newsletter in which he summarises his findings in the music industry. To gather information, he created a text document with a list of music resources that he checks regularly. Unfortunately, limited searches can be done on the document. Therefore, he would like to have access to an online catalogue that allows more sophisticated filtering options.
3. **Competency Questions.** The following questions were extracted from the story:
    a. CQ1: Can I search for musical content by applying filters (genre, period ...)?
    b. CQ2: What types of resources can I find?
    c. CQ3: Is the music resource X complete or incomplete?
    d. CQ4: Is a dataset attached to resource X?
    e. CQ5: Can I add resources as a user?
    f. CQ6: How can I share what I find on the site?
   Some preliminary requirements can already be detected from the questions, namely:
   ● Data requirements: genre and time (CQ1), type (CQ2) and availability (CQ4).
   ● Functional requirements: filters (CQ1), completeness (CQ3), crowdsourcing (CQ5), share (CQ6).
4. **Driver questions.** We identify CQ1 as the driving question, as it defines the problem space (*can I search for*), identifies the reference entity (*musical content*), and suggests how to visualise the data (*filters*).
5. **Clusters of questions.** Other CQs can be split in two groups: CQ2-4 address context information of the main entity (*type, complete/incomplete, dataset*); CQ5-6 address actions to be performed on the data (*add, share*).
6. **Select Data views.** The driver question matches a specific type of data visualisation, i.e. a filter-based exploration of resources ordered by relevance, hence no further analysis is needed.
7. **Interaction patterns**. In order to develop generous interfaces, we review web applications that present similar tasks, and then we map UI patterns to CQs:
    a. CQ1: group resources under categories and show the counting for each category, to give an overview.
    b. CQ2-CQ4: show lists of resources for each category on demand.
    c. CQ5, CQ6: provide specialised operations when browsing the record of a resource.
8. **Deploy solutions**. We check whether there are other personas with similar information requirements (step 1). Since no other personas have similar requests and the call for action is rather specific, we continue with the development of a bespoke solution, i.e. musoW.
9. **User test.** musoW was validated in focus groups with stakeholders, competitors, and project partners.

Laurent is the only persona that required a dedicated, specialised application for browsing music resources on the web. Other personas are either scholars with very specific research questions (for which we developed MELODY and the Corpus) or lay users, who do not have a specific task guiding their exploration (for which we develop a web portal). Iteratively applying the workflow to each persona may be time-consuming and does not ensure results are representative of the whole picture (rather, the result is simply going to be the sum of all requirements). Therefore, when analysing the remaining 18 personas, step 6 (Select Data views) is extended with a distant reading approach, performing an exploratory analysis of CQs. We manually annotated CQs in an online table[6] (see an example in Figure 2) with scope, classes, ontology patterns, and expected type of result (e.g. list, map, single result, explanation).

| CQ ID | exp. | CQ | Bibl. data | Music data | Linguis | User da | Main entities | Additional entities/props |
|---|---|---|---|---|---|---|---|---|
| Carolina1-CQ1 | - | Where was a musical composition performed? | yes | no | no | no | Musical performance: | musical composition |
| Carolina1-CQ2 | - | In which buildings was a musical composition performed? | yes | no | no | no | Musical perform | Musical performance: bui |
| Carolina1-CQ3 | - | Where was a musical composition performed for the first time? | yes | no | no | no | Musical perform | Musical performance: dat |

Figure 2. Manual annotation of Competency Questions

We analysed results to grasp an overview of priorities, data patterns, and user journeys. The preliminary analysis of CQs is available online as a Jupyter notebook[7]. In detail, we identified three categories of data and estimated their coverage:
● bibliographic data (on music works, historical events, composers) covered by 70% of CQs;
● structured music data (melody, harmony, rhythm), covered by 34% of CQs;
● linguistic, full-text data (emotions, song lyrics), covered by 30% of CQs. Among these, 77% rely also on bibliographic data, and 19% also on music data. Only 5% of CQs require all three types of data.

---
[6] https://docs.google.com/spreadsheets/d/16hr2fFTc4VUQHob0ALTu1TJ95xtyGvmWvly2yxOZFcM
[7] https://colab.research.google.com/drive/17I_3yjo2XoDDw6OTLvtAOE1PjOZ5xgmg?usp=sharing

We assume we can identify priorities as the most representative data requirements (bibliographic) and estimate the complexity of services to be implemented as those that satisfy niche areas (musicologists and linguists). Secondly, we analysed entities and ontology patterns. To this end, we identified an input, intermediate, and output entities for each CQ and we used a Sankey diagram to visualise journeys (Figure 3). For instance Carolina-CQ1 (Figure 2) "Where was a musical work performed" has "Music Work" as input, "Musical performance" as intermediate and "Place" as output.

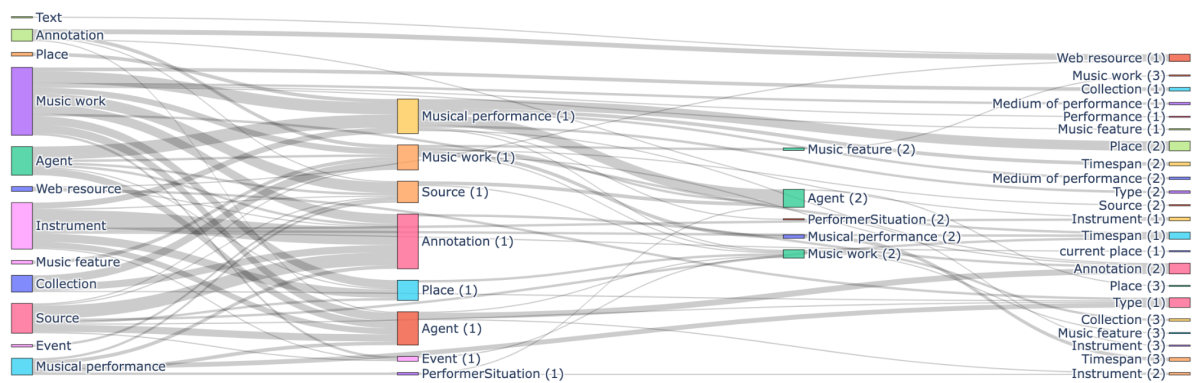

Figure 3. Data patterns and user journeys

Music works, agents, and sources are the main access points to the knowledge graph, intermediate aspects address more or less technical aspects, such as annotations (music and linguistic data), agents, and events, and final outputs are rather diverse. Again, we assume that most recurring patterns are priorities for developing UI components and journeys.

Step 7 (Interaction patterns) is performed via a competitive analysis, i.e. a user study wherein participants were requested to validate two similar web applications leveraging different UX strategies more/less similar to generous interfaces[8]. Results of the survey drove the final definition of UI components and their composition in the web page.

Finally, step 9 (User validation) is performed via another user study, this time devoted to co-design aspects[9]. In particular, users are asked to answer questions on how they would like or expect a website for music data recommendation to look like and behave. Users are asked to imagine themselves in a scenario they are comfortable with (e.g. "you are at home and you want to discover new music") and to describe their research process and expectations. It's worth noting that they do not see a website to evaluate. Results of the survey are matched against decisions already taken in step 7, which provide us, in a reverse-engineering fashion, with an evaluation of the expected user satisfaction. We believe it is fair to assume that the validation of the results produced by using our methodology can be inherited by the methodology itself.

# 5. CONCLUSION

We defined a workflow that leverages ontology requirements in UI/UX design to (1) develop ideas and prototypes that match data requirements, (2) have the UI/UX design iteratively informing and revising ontology requirements. The usage of exploratory analysis on Competency Questions gave us an overview of data requirements and a clear definition of priorities. Data patterns allow us to estimate types of content interaction and their relevance. Two user studies (focused respectively on competitive analysis and co-design techniques) help us to calibrate services and expectations of a wide range of users, including experts and lay people. Our preliminary results lead us to justify two strong assumptions, namely: (1) similar CQs can be grouped by type of interaction pattern; (2) entities that are relevant to a large number of CQs are also likely to be relevant to a wide range of users. As a consequence, we can apply our workflow to a much smaller number of groupings of CQs (instead of over each persona), therefore preventing time-consuming activities.

In conclusion, binding ontology design requirements to UI/UX design choices revealed being a good solution to tackle common issues in projects dedicated to the dissemination of cultural heritage data. Results on the case study were successful, and preliminarily validated the goodness of our approach, which was applied when designing four applications having different goals (a catalogue, an authoring platform, a linguistic corpus interface, and a web portal for engaging lay users). In future works we plan to test our methodology in projects with a different scope, in order to validate the reusability of our methodology in contexts different from the one in which it has been developed.

---

[8] https://docs.google.com/spreadsheets/d/1Q1Byjk9oAutD9yDk2Gaig-5VYqrSbRQ2SCu8Lz9MJf0
[9] https://docs.google.com/spreadsheets/d/1Ky83VoAUJRcMYSLlgaedw5QYH_s5KCNdPicUyWFq320

# 6. ACKNOWLEDGMENTS

This work is supported by a project that has received funding from the European Union's Horizon 2020 research and innovation programme under grant agreement No 101004746 (Polifonia: a digital harmoniser for musical heritage knowledge, H2020-SC6-TRANSFORMATIONS).